\newcommand{\be}{\begin{equation}}\newcommand{\ee}{\end{equation}}
\newcommand{\bea}{\begin{eqnarray}}\newcommand{\eea}{\end{eqnarray}}
\newcommand{\nn}{\nonumber \\}\newcommand{\p}[1]{(\ref{#1})}
\newcommand\T{\theta_{12}}

\newcommand\D{{\cal D}}

\newcommand{\xb}{\overline{\xi}}
\newcommand{\eb}{\overline{\eta}}
\newcommand{\GB}{\overline{G}}
\documentstyle[12pt]{article}

\begin{document}
\setcounter{page}0
\renewcommand{\thefootnote}{\fnsymbol{footnote}}
\thispagestyle{empty}
\begin{flushright}
LNF-97/028 (P)\\
JINR E2-97-141\\
hep-th/9707240 \\
July, 1997
\end{flushright}
\begin{center}
{\large\bf NONLINEAR REALIZATIONS OF SUPERCONFORMAL \\  
AND $W$ ALGEBRAS AS EMBEDDINGS OF STRINGS}
\vspace{0.5cm} \\
S. Bellucci\footnote{E-mail: bellucci@lnf.infn.it} \vspace{0.5cm} \\
{\it INFN - Laboratori Nazionali di Frascati} \\
{\it P.O.Box 13, I-00044 Frascati, Italy} \vspace{1cm}\\
V. Gribanov\footnote{E-mail: gribanov@thsun1.jinr.dubna.su}, 
E. Ivanov\footnote{E-mail: eivanov@thsun1.jinr.dubna.su},
S. Krivonos\footnote{E-mail: krivonos@thsun1.jinr.dubna.su}\\
and A. Pashnev\footnote{E-mail: pashnev@thsun1.jinr.dubna.su}
\vspace{0.5cm} \\
{\it JINR--Bogoliubov Laboratory of Theoretical Physics,         \\
141980 Dubna, Moscow Region, Russia} \vspace{1.5cm} \\
{\bf Abstract}
\end{center}
We propose a simple method for constructing representations of
(super)conformal and nonlinear $W$-type algebras in terms of their 
subalgebras and corresponding Nambu-Goldstone fields. We apply it to 
$N=2$ and $N=1$ superconformal algebras and describe in this way 
various embeddings of strings and superstrings for which these 
algebras and their subalgebras define world-sheet symmetries. 
Besides reproducing the known examples, we present some new 
ones, in particular an embedding of the bosonic string with  
additional $U(1)$ affine symmetry into $N=2$ superstring. We also apply 
our method to the nonlinear $W_3^{(2)}$ algebra and demonstrate that 
the linearization procedure worked out for it some time ago 
gets a natural interpretation as a kind of string embedding.
All these embeddings include the critical ones as particular 
cases.

\newpage
\renewcommand{\thefootnote}{\arabic{footnote}}
\setcounter{footnote}0
\setcounter{equation}0\section{Introduction}

For the last years, the study of various
embeddings of strings and superstrings received much 
attention \cite{BV}-\cite{BO}. This activity was initiated by 
the paper of Berkovits and Vafa \cite{BV} who showed that ordinary
bosonic strings can be regarded as a special class of vacua of $N=1$
superstrings. Later it was found that this is a general 
phenomenon: a (super)string with $N$ extended world-sheet supersymmetry 
can be embedded into a superstring with $N+1$ extended supersymmetry 
as a particular vacuum state of the latter. Analogous embeddings were 
constructed for strings associated with nonlinear $W$ type  
algebras and their linearizing algebras \cite{boh}-\cite{bbr}.
It was suggested that 
all the known strings and superstrings present different vacua of some 
hypothetical universal string theory. 

An essential step towards clarifying the group-theoretical grounds of 
the embedding procedure was made in refs. \cite{{K},{McA}}. 
On the example of 
the bosonic string embedded into the $N=1$ superstring 
Kunitomo \cite{K} showed 
that the larger symmetry ($N=1$ superconformal) is realized with 
the help of Nambu-Goldstone fields, typical for the spontaneously broken 
supersymmetry. The vacuum stability subgroup is the Virasoro one, just 
the world-sheet symmetry of the bosonic string. The same results have 
been obtained by McArthur \cite{McA} in the framework of the standard 
theory of nonlinear realizations \cite{CWZ,V} applied to the $N=1$
superconformal 
algebra (SCA). These observations support a nice interpretation of 
the string embeddings as one more manifestation of the universal 
phenomenon of spontaneous symmetry breakdown, this time of the
infinite-dimensional world-sheet (super)symmetry of strings \cite{BV}. 
 From this point of view, any embedding of the lower-symmetry string 
into the 
larger-symmetry one amounts to the choice of special background (and hence 
the vacuum) for the latter, such that it is given by the Nambu-Goldstone 
fields realizing the spontaneous breaking of the larger symmetry down to the 
lower symmetry. The currents generating the larger symmetry are 
expressed in terms of those of the lower symmetry and the Nambu-Goldstone
fields. A power of the nonlinear realizations method manifests itself
in that these 
expressions can be obtained in an algorithmic way, knowing only the 
structure relations of the given superconformal algebra. In accord 
with the general concepts of this method, all the basic characteristics 
of such specific representation of the larger symmetry 
(central charges, etc.) should be fully determined by the structure of 
representations of the vacuum stability 
symmetry. In this way, the string theory associated with the lower symmetry 
comes out as a spontaneously broken phase of the higher-symmetry 
string theory. 
   
In this paper we present simple and universal techniques of 
calculations for nonlinear realizations of infinite dimensional algebras 
written in terms of OPEs (SOPEs) for one dimensional currents 
(supercurrents). The 
application of this techniques
to $N=1,\;2$ superconformal algebras after taking into account quantum 
corrections leads to the corresponding formulas
for the string embeddings. 

Besides reproducing the known $N=1 \rightarrow 
N=2$ embedding in terms of $N=1$ superfields \cite{BO}, we get new 
self-consistent embeddings 
by choosing different subalgebras of $N=2$ SCA as 
the vacuum stability symmetries. In particular, we describe embedding 
of the string associated with the product of Virasoro and $U(1)$ Kac-Moody 
algebras into $N=2$ superstring. This extension of the bosonic string 
was recently discussed \cite{{BK},{far}} in connection with 
$F$-theory \cite{Vafa}. As a by-product, we also reproduce the  
$N=0 \rightarrow N=1$ embeddings within our techniques, 
discuss how they are related to the $N=2$ embeddings 
constructed and make comparison with the results of 
refs. \cite{{K},{McA}}. 

We argue that the linearization 
procedure for some $W$ algebras worked out 
some time ago \cite{KS} can also be interpreted in the embedding 
language, namely as an embedding of the string associated with some 
linear subalgebra of the given nonlinear $W$ (super)algebra into 
the string associated with this $W$ (super)algebra itself. The appropriately 
modified nonlinear realizations techniques prove to work in 
this case too. We consider the simple example of the quasi-conformal 
algebra $W_3^{(2)}$, but the same is apparently true for a wider 
class of $W$ algebras. 

In Sect. 2 we give a general characterization of our method. In Sect. 3 
we apply it to $N=2$ SCA and describe embeddings of strings associated with 
various subalgebras of $N=2$ SCA into the $N=2$ superstring. Sect. 4 is 
devoted to applications of our method to nonlinear algebras 
on the example of the $W_3^{(2)}$ algebra.

\setcounter{equation}0
\section{Method of nonlinear realizations}
As a starting point, let us briefly describe the theory of nonlinear 
realizations \cite{CWZ,V} with some refinement related to the 
case of infinite-dimensional 
symmetries. This theory is a set of recipes of how to realize the 
given group $G$ on the parameters of its coset $G/H$, $H$ being some 
subgroup of $G$.     

After choosing the subgroup $H$, one represents an arbitrary group 
element in the
exponential parametrization in the following form
\begin{equation}
G=K H=e^k e^h \;.
\end{equation}
Here $h= \xi^ah_a$ and $k = \phi^ik_i$ belong, respectively, to the 
algebra of $H$ and to its complement to the full algebra of $G$, $h_a$ and 
$k_i$ being the appropriate generators. An 
infinitesimal group element $g=1+\epsilon$, with $\epsilon = 
\epsilon^ik_i + \epsilon^ah_a$, has the following action on $K$
\begin{equation}\label{1}
(1+\epsilon)e^k=e^{k+\delta k}(1+\delta h)\;,
\end{equation}
whence 
\begin{equation}
e^{-k}\epsilon e^k=e^{-k}\delta e^k+\delta h \;.
\end{equation}
Both the left- and right-hand sides of this equation can be written 
in terms of multiple commutators
\begin{equation} \label{2}
e^{-k} \wedge \epsilon=\frac{1-e^{-k}}{k} \wedge \delta k + \delta h\;.
\end{equation}
This is the basic equation for determining $\delta k$ and $\delta h$.
The definition of the symbol $\wedge$ is as follows. For the given
function $f(k)=f_0+f_1 k+f_2 k^2+...$ it reads:
\be
f(k)\wedge \epsilon=\epsilon+f_1[k,\epsilon]+
f_2[k[k,\epsilon]]+... \;.
\ee
For the coefficients of generators in 
$\delta k=\delta \phi^ik_i,\;\delta h=\delta \xi^ah_a$ eq. \p{2} 
implies the following general expressions
\begin{eqnarray}
\delta \phi^l&=&\epsilon^iF_i^l+\epsilon^aF_a^l \;,\nn
\delta \xi^b &=&\epsilon^iF_i^b+\epsilon^aF_a^b\;.
\end{eqnarray}
Here all $F$'s are some functions of $\phi^k$. They are uniquely specified 
by the structure relations of the $G$ algebra.
Further, one considers a space of functions $\Phi(\phi^i)$ 
on which some representation of the subalgebra $h$ is realized 
(it is reducible in general). In 
what follows we will call it the `matter' representation and 
denote its generators by  $h_a^{(m)}$.
Then the generators of the whole algebra can be  
realized on this space as
\cite{McA}:
\begin{eqnarray}
H_a&=&-F_a^l\frac{\partial}{\partial\phi^l}+F_a^bh_b^{(m)}\;,\nn \label{r2}
K_i&=&-F_i^l\frac{\partial}{\partial\phi^l}+F_i^bh_b^{(m)}\;.
\end{eqnarray}
To prevent a possible confusion, we point out that the original generators 
(those appearing in eqs. (2.1) - (2.6)) are always assumed to be 
abstract and 
commuting with all coset parameters, it is the algebra of their commutators 
which really enters the game. On the contrary, the generators \p{r2} give 
a particular realization of the group $G$ on the space of 
these parameters and functions of them.  

In the case of infinite-dimensional algebras, such as the superconformal
ones, it is more convenient to deal with currents or 
supercurrents like  
$T(Z)=T(z,\theta^a)$,
and with OPEs or SOPEs instead of the commutation relations. 
A finite number of such (super)currents collects all the infinite 
set of generators of the given (super)algebra. These 
generators appear as coefficients  in the
$\theta$ and $z$  expansions of supercurrents 
(as usual, Laurent series is assumed for the $z$ expansion). 
An element of the algebra 
associated with the given current is expressed as an 
integral over the superspace with some parameter
depending on $z,\theta^a$:
\begin{equation} \label{integ}
{\cal T}=\frac{1}{2\pi i}\oint dZ \phi(Z) T(Z) \;.
\end{equation}
The coset parameters $\phi^m$ come out now as the 
coefficients in the expansion of $\phi(Z)$ over $Z$.
As before, it is assumed that the original abstract currents 
(supercurrents) have vanishing OPEs (SOPEs) with all $\phi(z)$.

After computing group variations of these parameters-functions, 
the representations
similar to \p{r2} arise. An essentially new 
point compared to the customary nonlinear realizations formalism is 
the necessity to introduce new (super)currents $p(Z)$ 
which substitute  the derivatives
$\frac{\partial}{\partial\phi^l}$. These new  currents are canonically
conjugated to $\phi (Z)$. This means that the following OPE
\begin{equation}
p_l(z)\phi^n(w)\sim\delta_l^n\frac{1}{z-w}\;,
\end{equation}
or its obvious supersymmetric counterpart are valid.
The resulting specific representation for the currents of the algebra reads
\begin{eqnarray}
H_a(Z)&=&-F_a^mp_l(Z)+F_a^bh_b^{(m)}(Z)\;,\nn \label{x2}
K_i(Z)&=&-F_i^mp_m(Z)+F_i^bh_b^{(m)}(Z)\;.
\end{eqnarray}
All $F$'s are now functions of $\phi^l(Z)$ and their derivatives.
The indices $a, b$ label the currents generating some infinite-
dimensional subalgebra of the given algebra (the stability subalgebra) 
while the indices $i,k,l$ refer to the remainder of currents (the coset 
ones). The `matter' currents $h^{(m)}_b$ still have vanishing OPEs 
with the coset parameters $\phi_k$ and conjugated momenta $p_l$. 
At the same time, OPEs between $h^{(m)}_b$ form a representation 
of the stability subalgebra. 
The action of \p{x2} on the relevant functionals of $\phi(Z)$ (analogs of 
$\Phi(\phi^k)$) is defined through setting OPEs between these objects.
The transformation properties of these functionals with respect to the 
whole (super)algebra are fully determined by their transformation properties 
with respect to the stability subalgebra, i.e. by fixing their OPEs with 
the `matter' (super)currents $h^{(m)}_b$ (these functionals are a sort of 
primary (super)fileds of the stability subalgebra).

We emphasize that all the 
infinite-dimensional nature of the algebra and the stability 
subalgebra is hidden in the 
integrals like \p{integ}, while the explicitly appearing indices run over  
finite ranges like in the case of finite-dimensional symmetries. 
      
One essential remark at this stage is of need. The currents \p{x2}
realize OPEs of the initial algebra only if we consider them as classical
ones, i.e. when we keep in OPEs only single contraction. This is
equivalent to using the Poisson brackets and implies that the
subalgebras generated by $h_a^{(m)}(Z)$ and $H_a(Z)$ are identical. 
In particular,
they have the same central charges. Transition to the exact quantum OPEs 
by keeping
all contractions radically changes the situation. OPEs will 
not close unless some terms (quantum corrections) are 
added to the classical expressions for the currents. 
It can be shown that these addings do not include $p_i$'s. 
In this case central charges in the subalgebras generated  by $h_a^{(m)}$ 
and $H_a$ become different
and the above formalism gets suitable for description of embeddings 
of (super)strings along the lines of refs. \cite{BV}-\cite{bbr}.

This method, as it was described above, suits very well the case of
linear algebras. Some additional problems arise when the $W$ type nonlinear 
algebras are regarded. Nevertheless, it seems to be still applicable 
at least when some {\it linear} subalgebra of given $W$ algebra is chosen 
as the vacuum stability subalgebra. In Sec. 4 we demonstrate this on 
the example of $W_3^{(2)}$ algebra.

\setcounter{equation}0

\section{Nonlinear realizations of $N=2$ SCA}
\subsection{Structure relations}

We start with some definitions. OPEs for $N=2$ SCA in terms of real
currents are given by 
\begin{eqnarray} \label{algebra}
T(z)T(w)&\sim&\frac{c/2}{(z-w)^4}+\frac{2T(w)}{(z-w)^2}+
\frac{T'(w)}{z-w}\; ,\;
T(z)J(w) \sim \frac{J(w)}{(z-w)^2}+
\frac{J'(w)}{z-w}\;, \nn
T(z)G_{1,2}(w)&\sim&\frac{3/2G_{1,2}(w)}{(z-w)^2}+
\frac{G'_{1,2}(w)}{z-w}\; , \;
G_{1,2}(z)G_{1,2}(w) \sim \frac{c/3}{(z-w)^3}+
\frac{T(w)}{z-w}\; ,\nn
G_{1}(z)G_{2}(w)&\sim&\frac{J(w)}{(z-w)^2}+
\frac{1/2 J'(w)}{z-w}\; ,\;
J(z)J(w) \sim -\frac{c/3}{(z-w)^2} \;, \nn
J(z)G_{1}(w)&\sim&-\frac{G_{2}(w)}{(z-w)}\; ,\;
J(z)G_{2}(w) \sim \frac{G_{1}(w)}{(z-w)}\;. \label{n2sca}
\end{eqnarray}
Note that the supercurrents $\{T,G_1\}$ and $\{T,G_2\}$ form two different 
$N=1$ SCAs 
embedded as subalgebras into the $N=2$ SCA.

In terms of $N=1$ supercurrents $T(Z)=\frac{1}{\sqrt{2}}G_1(z)+\theta T(z)$
and $G(Z)=-i\bigl (J(z)+\theta\sqrt{2} G_2(z)\bigr )\;\;$ these OPEs 
are concisely presented by the following SOPEs
\begin{eqnarray}
T(Z_1)T(Z_2)&\sim&\frac{c/6}{Z_{12}^3}+\frac{3/2\T T(Z_2)}{Z_{12}^2}+
\frac{1/2\D T(Z_2)+\T T'(Z_2)}{Z_{12}}\;,\nn
T(Z_1)G(Z_2)&\sim&\frac{\T G(Z_2)}{Z_{12}^2}+
\frac{1/2\D G(Z_2)+\T G'(Z_2)}{Z_{12}}\;,\nn
G(Z_1)G(Z_2)&\sim&\frac{c/3}{Z_{12}^2}+\frac{2\T T(Z_2)}{Z_{12}}\;,
\end{eqnarray}
where
\begin{equation}
\T =\theta_1-\theta_2,\; Z_{12}=z_1-z_2-\theta_1\theta_2,\;
\D=\frac{\partial}{\partial\theta}+\theta\partial,\;T'\equiv\frac{
\partial T}{\partial z}.
\end{equation}

In the rest of this Section we apply the general method described 
in Sect. 2 in order to 
construct nonlinear realizations of $N=2$ SCA with its different 
subalgebras as the vacuum stability subalgebra $H$. For different choices 
of $H$ the coset is parametrized by different sets of fields, which 
gives rise to non-equivalent realizations of $N=2$ SCA. The quantum 
versions of the latter are easy to construct, and they give the 
formulas for the corresponding embeddings. We also show that 
the nonlinear realizations of $N=1$ SCA and the associated  
$N=0 \rightarrow N=1$ embeddings follow from the $N=2$ ones upon  
appropriate reductions. 
 
\subsection{Stability subalgebra $H = \{T(Z)\}=\{T(z), G_1(z) \}$} 
We first choose $H$ to be generated by currents  $T(z), G_1(z)$, or, in the 
$N=1$ superfield notation, by the spin $3/2$ fermionic supercurrent $T(Z)$. 
This is just $N=1$ SCA. 

The general element
of the coset is parametrized as
\begin{equation}
K=e^{\frac{1}{2\pi i}\oint dZ \phi(Z) G(Z)}\;,
\end{equation}
where we have introduced a Nambu-Goldstone fermionic $N=1$ superfield 
$\phi(Z)$ with the ``spin'' $-1/2$. 
An infinitesimal element of the $N=2$  superconformal 
group
\begin{equation}
g=1+\frac{1}{2\pi i}\oint dZ \{\alpha(Z) G(Z)+a(Z) T(Z)\}\;,
\end{equation}
acts on $K$ according to the following relation (cf. \p{1})
\begin{equation}
g e^{\frac{1}{2\pi i}\oint dZ \phi(Z) G(Z)}=
e^{\frac{1}{2\pi i}\oint dZ (\phi(Z)+\delta\phi(Z)) G(Z)}\tilde H.
\end{equation}
Under this left shift the parameter of the coset space $\phi(Z)$ changes 
to $\phi(Z)+\delta\phi(Z)$. The induced subgroup element $\tilde{H}$ 
is represented by 
\begin{equation}\label{h}
\tilde{H}=1+\frac{1}{2\pi i}\oint dZ \{b(Z) T(Z)+c(Z))\}.
\end{equation}
The quantities $\delta \phi(Z)$, $b(Z)$ and $c(Z)$ are to be expressed 
in terms of the Nambu-Goldstone superfield $\phi(Z)$ and the 
group parameters $a(Z)$ and $\alpha(Z)$. 

While applying the techniques described in the previous 
Section, one frequently needs to compute 
(anti)commutators of the operators like $\frac{1}{2\pi i}\oint dZ
 \phi(Z) G(Z)$. This can be done with the help of the following
formula:
\begin{eqnarray}\label{com}
\left[\frac{1}{2\pi i}\oint dZ_1  a(Z_1) A(Z_1),
\frac{1}{2\pi i}\oint dZ_2 b(Z_2) B(Z_2)\right]&=&\nn
\sigma\left(\frac{1}{2\pi i} \right)^2\oint dZ_2 \oint_{C} dZ_1 
 a(Z_1) b(Z_2) A(Z_1)B(Z_2)\;.&&
\end{eqnarray}
The contour of integration $C$ surrounds $z_2$ and $A(Z_1)B(Z_2)$
in the right-hand side of \p{com} stands for SOPE of the 
supercurrents $A(Z_1)$ and $B(Z_2)$.
Additional multiplier $\sigma$ is the sign which depends on 
the Grassmann parity
of $A(Z)$ and $B(Z)$: $\sigma=(-1)^{g(A)(g(B)+ 1)}$.

Taking this into account and doing the computations along the line 
of Sec. 2, we find the following expressions for 
$\delta\phi(Z)$, $b(Z)$ and $c(Z)$ 
\begin{eqnarray}
\delta_a\phi&=& \frac{1}{2}a'\phi-a\phi'-\frac{1}{2}\D a\D\phi\;,\nn
\delta_{\alpha}\phi&=&\alpha\left\{\D\phi\coth\D\phi+
\frac{\phi\phi'}{(\D\phi)^2}
\left(1-2\D\phi\coth\D\phi+\frac{(\D\phi)^2}{\sinh^2\D\phi}
\right)\right\}\nn 
&&+ \frac{\D\alpha\phi}{\D\phi}\left(1-\D\phi\coth\D\phi\right)\;,\nn 
b(Z)&=&a+\alpha\frac{2\phi}{\D\phi}\tanh\left(\frac{\D\phi}{2}\right)\;,\nn 
c(Z)&=&-\alpha\frac{c_m}{6}\left\{\D\phi'+ 
\left(\frac{2\phi\phi'\D\phi'}{(\D\phi)^3}- 
\frac{\phi\phi''}{(\D\phi)^2}\right) 
\left(\D\phi-2\tanh\left(\frac{\D\phi}{2}\right)\right) \right\}\;.
\end{eqnarray}
These expressions lead to the following form of the generators:
\begin{eqnarray}
T&=&-\phi'\eta-\frac{1}{2}\phi\eta'+\frac{1}{2}\D\phi\D\eta+T_m\;,\nn
G&=&\eta+\frac{2\phi\phi'}{(\D\phi)^2}\left(1-\D\phi\coth\D\phi\right)\eta
-\frac{\phi}{\D\phi}\left(1-\D\phi\coth\D\phi\right)\D\eta
\nn
&& -\frac{c_m}{6}\left\{\D\phi'+
\left(\frac{2\phi\phi'\D\phi'}{(\D\phi)^3}- 
\frac{\phi\phi''}{(\D\phi)^2}\right) 
\left(\D\phi-2\tanh\left(\frac{\D\phi}{2}\right)\right)\right\} \nn
&& +\frac{2\phi}{\D\phi}\tanh\left(\frac{\D\phi}{2}\right)T_m\;,  
\end{eqnarray}
where the newly introduced spin $1$ bosonic superfield $\eta$ is canonically 
conjugated to $\phi$
\begin{equation}
\eta(Z_1)\phi(Z_2)\sim\frac{\T}{z_{12}}\;.
\end{equation}
Thus we have obtained the realization of the $N=2$ SCA in terms of the 
conjugated pair of $N=1$ Nambu-Goldstone superfields $\phi$, $\eta$ and 
$N=1$ SCA supercurrent $T_m(Z)$
\begin{eqnarray}
T_m(Z_1)T_m(Z_2) & \sim & \frac{c_m/6}{Z_{12}^3}+
\frac{3/2\T T_m(Z_2)}{Z_{12}^2}+
\frac{1/2\D T_m(Z_2)+\T T_m'(Z_2)}{Z_{12}}\;, \nn
T_m(Z_1)\phi(Z_2) & \sim & 0 \; , \;
T_m(Z_1)\eta(Z_2)  \sim 0 \;. \nonumber
\eea
At the considered classical level the 
central charge of $N=2$ SCA in such a realization can be checked to 
coincide with the central charge of $N=1$ SCA
$$
c_{N=2} = c_m\;.
$$

As was already mentioned in Introduction, the main motive  
for working out this approach to nonlinear realizations of 
(super)conformal (and $W$) algebras was the desire to gain a 
systematic method for deriving the relations which describe different 
embeddings of strings. According to the reasoning of refs. \cite{K,McA}, 
the linearly realized subgroup always corresponds to the embedded 
string and defines the world-sheet symmetry of the latter. Thus  
in the present case we should get an embedding of $N=1$ superstring into 
$N=2$ superstring. This embedding in terms of $N=1$ superfields was firstly 
given by Berkovits and Ohta \cite{BO} by means of guesswork. Now 
we show that their formulas naturally follow from the above ones 
obtained within a systematic procedure, with taking account of 
quantum corrections.

To make a comparison with the Berkovits-Ohta paper 
\cite{BO}, we perform  the canonical
transformation from the superfields $\phi,\;\eta$ to the superfields 
$C,\;B$
\begin{eqnarray}
C&=&
\frac{2\phi}{\D\phi}\tanh(\frac{\D\phi}{2})\;,\nn
B&=&\frac{1}{2}(1+\cosh\D\phi)\eta+\frac{\cosh^2(\frac{\D\phi}{2})}{
(\D\phi)^2}\left(2-\D\phi\coth\left(\frac{\D\phi}{2}\right)\right)
\phi\phi'\eta \nn
&&+ \frac{\phi}{2\D\phi}\left(\D\phi\coth\left(\frac{\D\phi}{2}\right)-
\cosh\D\phi-
1\right)\D\eta\;. 
\end{eqnarray}
The generators of $N=2$ SCA in terms of these new superfields take 
the form:
\begin{eqnarray}
T&=&-C'B-\frac{1}{2}CB'+\frac{1}{2}\D{C}\D{B}+T_m\;, \nn
G&=&B-\frac{1}{4}(\D{C})^2B+\frac{1}{2}BC'C+
\frac{1}{2}\D{C}C\D{B}+CT_m \nn
&&- \frac{c_m}{6}\left(\frac{4\D{C}'}{4-(\D{C})^2}-
\frac{2CC'\D{C}'(\D{C})^2}{(4-(\D{C})^2)^2}-
\frac{CC''\D{C}}{4-(\D{C})^2}\right),\label{G}
\end{eqnarray}
and almost coincide with those of \cite{BO}. The only difference is  
the presence of some additional $B$-independent 
terms in the realization of ref. \cite{BO}. The origin of this difference 
lies in 
the following. 
SOPEs for the supercurrents
 \p{G} are closed on the classical level, when only one contraction
is taken into account. Besides, the values of the 
central charge for $T_m$ and $T$ on the classical level are the same.
The extra terms found in ref. \cite{BO}
can be easily restored in our formulation by demanding the closure 
of the {\it quantum}
SOPEs, when all contractions are taken into
account. The final expressions for the quantum supercurrents are \cite{BO}:
\begin{eqnarray}
T^q&=&T+\left( \frac{C\D{C'}}{4-(\D{C})^2}\right)',\nn
G^q&=&G+\frac{C C'\D{C}'}{4-(\D{C})^2}\;.
\end{eqnarray}
As was stated in \cite{BO}, these expressions describe both the cases of 
critical and non-critical embeddings. It is a matter of direct 
computation to see that the central charge $c_{N=2}$ in this quantum
realization is related to the central charge $c_m$ inherent to $T_m$ as 
\be 
c_{N=2} = c_m - 9\;. 
\ee
The critical value of $c_m = 15$ yields just the critical value 
$c_{N=2}=6$ for the central charge of $N=2$ SCA.  

\subsection{Stability subalgebra $H=\{T(z)\}$}
In our second example 
we take as the linearly realized subalgebra the Virasoro subalgebra. 
So it should give, after passing to the quantum case, 
the description of embedding of bosonic string into the $N=2$ superstring.  
As distinct from the previous example, all supersymmetries are 
nonlinearly realized in the case under consideration. As a result, no 
superfield formalism exists, and one should deal with the component currents 
rather than supercurrents.

We are led to introduce three Nambu-Goldstone fields 
associated with the currents $$
\GB (z)=\frac{1}{\sqrt{2}}\bigl( G_1(z)-iG_2(z)\bigr)\;,\;
{G}(z)=-\frac{1}{\sqrt{2}}\bigl( G_1(z)+iG_2(z)\bigr)\;\mbox{ and }\;
\stackrel{\sim}{J}(z)=-iJ(z)\;.$$
We can parametrize the coset in two different ways, each leading to
different representations of the algebra in 
terms of $T_m(z)$, Nambu-Goldstone fields
$\xi(z), \overline{\xi}(z), \phi(z)$ and their conjugated momenta
$\overline{\eta}(z), \eta(z), \mu(z)$ with the following OPEs:
\begin{equation}
\eta(z)\overline{\xi}(w)\sim\frac{1}{z-w}\;, \;
\overline{\eta}(z){\xi}(w)\sim\frac{1}{z-w}\;, \;
\mu(z){\phi}(w)\sim\frac{1}{z-w}\;.
\end{equation}
So far as the classical case is concerned, these different 
parametrizations are obviously 
related by an equivalence  
transformation (it can still be 
rather complicated), but in the quantum case they can yield non-equivalent  
realizations. This is the reason why we quote the latter for both 
parametrizations. 
 
Let us first consider the `symmetric' parametrization
\begin{equation}
K=e^{\frac{1}{2\pi i}\oint dz \{\overline{\xi}(z) G(z)+\xi(z)\overline{G}
(z)\}}e^{\frac{1}{2\pi i}\oint dz \phi(z) \stackrel{\sim}{J}(z)}\;.
\end{equation}
A straightforward computation leads to the following expressions
for the supercurrents:
\begin{eqnarray}                            
T&=&T_m+\frac{3}{2}\xi'\overline{\eta}+\frac{3}{2}\overline{\xi}'\eta+
\frac{1}{2}\xi\overline{\eta}'+\frac{1}{2}\overline{\xi}\eta'+\phi'\mu\;,\nn
\stackrel{\sim}{J}&=&-\mu-\overline{\xi}\eta+\xi
\overline{\eta}-\frac{c_m}{6}\phi'\;, \label{b2} \nn
G&=&-\eta+\frac{2}{3}\xi\overline{\xi}'\eta+
\frac{1}{3}\xi'\overline{\xi}\eta+\frac{1}{3}\xi\overline{\xi}\eta'-
\frac{1}{9}\xi\xi'\overline{\xi}'\overline{\xi}\eta+
\frac{2}{3}\xi\xi'\overline{\eta} \nn
&& + \frac{1}{2}(\xi+\frac{1}{6}\xi\xi'\overline{\xi})T_m+
\frac{1}{4}\{ (2\xi'+\frac{1}{6}\xi\overline{\xi}\xi''+
\frac{2}{3}\xi\overline{\xi}'\xi'+2\xi\phi'+\frac{1}{3}\xi\xi'
\overline{\xi}\phi')\mu\nn
&&+(\xi+\frac{1}{3}\xi\overline{\xi}\xi')\mu'\}+
\frac{c_m}{12}\{\xi''+\frac{1}{12}(2\xi\overline{\xi}'\xi''+
\xi'\overline{\xi}\xi''+\xi\overline{\xi}\xi'''+
2\xi\xi'\overline{\xi}'')
\nn 
&& + \frac{1}{24}\xi\xi'\overline{\xi}\overline{\xi}'\xi''+  
(\xi'+\frac{1}{12}\xi\overline{\xi}\xi''+
\frac{1}{3}\xi\overline{\xi}'\xi')\phi'+
\frac{1}{2}(\xi+\frac{1}{3}\xi\overline{\xi}\xi')\phi''\}\;,\nn
\overline{G}&=&-\overline{\eta}+
\frac{2}{3}\overline{\xi}\xi'\overline{\eta}+
\frac{1}{3}\overline{\xi}'{\xi}\overline{\eta}+
\frac{1}{3}\overline{\xi}{\xi}\overline{\eta}'-
\frac{1}{9}\overline{\xi}\overline{\xi}'{\xi}'{\xi}\overline{\eta}+
\frac{2}{3}\overline{\xi}\overline{\xi}'{\eta} \nn 
&&+ \frac{1}{2}(\overline{\xi}+\frac{1}{6}
\overline{\xi}\overline{\xi}'{\xi})T_m-
\frac{1}{4}\{ (2\overline{\xi}'+\frac{1}{6}\overline{\xi}{\xi}
\overline{\xi}''+
\frac{2}{3}\overline{\xi}{\xi}'\overline{\xi}'
-2\overline{\xi}\phi'-\frac{1}{3}\overline{\xi}\overline{\xi}'
{\xi}\phi')\mu
\nn
&&+ (\overline{\xi}+\frac{1}{3}\overline{\xi}{\xi}\overline{\xi}')\mu'\} +
\frac{c_m}{12}\{\overline{\xi}''+\frac{1}{12}(2\overline{\xi}{\xi}'
\overline{\xi}''+
\overline{\xi}'{\xi}\overline{\xi}''+\overline{\xi}{\xi}\overline{\xi}'''+
2\overline{\xi}\overline{\xi}'{\xi}'')
\nn
&&+\frac{1}{24}\overline{\xi}\overline{\xi}'{\xi}{\xi}'\overline{\xi}''- 
(\overline{\xi}'+\frac{1}{12}\overline{\xi}{\xi}\overline{\xi}''+
\frac{1}{3}\overline{\xi}{\xi}'\overline{\xi}')\phi'-
\frac{1}{2}(\overline{\xi}+
\frac{1}{3}\overline{\xi}{\xi}\overline{\xi}')\phi''\}\;.\label{b4}
\end{eqnarray}

In the case of the `non-symmetric' parametrization,
\begin{equation}
K=e^{\frac{1}{2\pi i}\oint dz \overline{\xi}(z) G(z)}
e^{\frac{1}{2\pi i}\oint dz \xi(z)\overline{G}(z)}
e^{\frac{1}{2\pi i}\oint dz \phi(z) \stackrel{\sim}{J}(z)}\;,
\end{equation}
the expressions for the currents of the algebra are much simpler:
\begin{eqnarray}
T&=&T_m+\frac{3}{2}\xi'\overline{\eta}+\frac{3}{2}\overline{\xi}'\eta+
\frac{1}{2}\xi\overline{\eta}'+\frac{1}{2}\overline{\xi}\eta'+\phi'\mu\;,
\nn
\stackrel{\sim}{J}&=&-\mu-\overline{\xi}\eta+\xi\overline{\eta}-
\frac{c_m}{6}
\phi'\;,  \nn
G&=&-\eta\;, \nn
\overline{G}&=&-\overline{\eta}+
2\overline{\xi}\xi'\overline{\eta}+
\overline{\xi}'{\xi}\overline{\eta}+
\overline{\xi}{\xi}\overline{\eta}'+
\overline{\xi}\overline{\xi}'\eta \nn 
&&+ \overline{\xi}T_m-\overline{\xi}'\mu-
\frac{1}{2}\overline{\xi}\mu'+\overline{\xi}\phi'\mu +
\frac{c_m}{12}(2\overline{\xi}''-
2\overline{\xi}'\phi'-\overline{\xi}\phi'')\;. \label{b8}
\end{eqnarray}

Once again, OPEs for the currents
\p{b4} and \p{b8} are closed only on the classical level
and with the same values of the central charge
for $T_m$ and $T$. The corresponding quantum expressions 
are
\begin{eqnarray}
T^q&=&T+\frac{1}{6}(\overline{\xi}''\xi'+\xi''\overline{\xi}'+
\overline{\xi}'''\xi+\xi'''\overline{\xi})+
\frac{1}{9}\overline{\xi}''\xi''\overline{\xi}\xi \nn
&&- \frac{1}{18}(\overline{\xi}'\xi'\overline{\xi}\xi''+
\xi'\overline{\xi}'\xi\overline{\xi}''+
\overline{\xi}'\xi\overline{\xi}\xi'''
+\xi'\overline{\xi}\xi\overline{\xi}''')\;,\nn
\tilde{J}^q&=&\tilde{J}+\frac{13}{3}\phi'-\frac{2}{3}\overline{\xi}'\xi'-
\frac13(\overline{\xi}''\xi-\xi''\overline{\xi})\;, \nn
G^q&=&G-\frac{5}{3}\xi''-\frac{3}{8}\xi\xi'\overline{\xi}''+
\frac{7}{24}\xi\xi''\overline{\xi}'-\frac{1}{72}\xi'\overline{\xi}\xi''+
\frac{23}{108}\xi\xi'''\overline{\xi}+
\frac{49}{432}\overline{\xi}'\xi'\overline{\xi}\xi\xi''
 \nn
&& +\frac{13}{18}\phi'\xi\xi'\overline{\xi}' +
\frac{13}{72}\phi'\xi\xi''\overline{\xi}+
\frac{13}{36}\phi''\xi\xi'\overline{\xi}-
\frac{13}{6}\phi'\xi'-\frac{13}{12}\phi''\xi\;,    \nn
\overline{G}^q&=&\overline{G}-\frac{5}{3}\overline{\xi}''-
\frac{3}{8}\overline{\xi}\overline{\xi}'\xi''+
\frac{7}{24}\overline{\xi}\overline{\xi}''\xi'-
\frac{1}{72}\overline{\xi}'\xi\overline{\xi}''+
\frac{23}{108}\overline{\xi}\overline{\xi}'''\xi+
\frac{49}{432}\xi'\overline{\xi}'\xi\overline{\xi}\overline{\xi}''
\nn 
&&-\frac{13}{18}\phi'\overline{\xi}\overline{\xi}'\xi' -
\frac{13}{72}\phi'\overline{\xi}\overline{\xi}''\xi-
\frac{13}{36}\phi''\overline{\xi}\overline{\xi}'\xi+
\frac{13}{6}\phi'\overline{\xi}'+\frac{13}{12}\phi''\overline{\xi} 
\label{bq4}
\end{eqnarray}
for the symmetric case and
\begin{eqnarray}
T^q&=&T\;,\;                 
\tilde{J}^q = \tilde{J}+\frac{13}{3}\phi'\;,  \nn
G^q&=&G\;,\;          
\overline{G}^q = \overline{G}-\frac{10}{3}\overline{\xi}''+
\frac{13}{3}\phi'\overline{\xi}'+\frac{13}{6}\phi''
\overline{\xi} \label{bq8}
\end{eqnarray}
for the non-symmetric one.
Expressions \p{b4}, \p{bq4} and  
\p{b8}, \p{bq8}  describe two possible embeddings 
of bosonic string into $N=2$ string. To our knowledge, they 
were not given before in literature. In both cases one readily checks 
the following relation between the central charges $c_{N=2}$ and $c_m$ 
\be  \label{crit0}
c_{N=2} = c_m - 20\;.
\ee
Once again, the critical value of the Virasoro central charge 
$c_m = 26$ yields the critical value $c_{N=2}=6$ for the central 
charge of $N=2$ SCA (it corresponds to the bosonic critical 
dimension $d_{N=2} = 4$). 

For completeness and for the sake of comparison with other options given 
below, let us remind the well-known form of BRST operator for 
the Virasoro algebra $H=\{ T_m^q \}$ 
\be  \label{brst4}
Q = \frac{1}{2\pi i}\oint dz \; c (T_m^q+\frac{1}{2}T_{gh})
\ee
with
\be
T_{gh} = cb'+2 c'b \; .
\ee 
Condition for the nilpotency of the BRST operator \p{brst4} is just 
$c_m=26$.

Note that the constructed embedding of $N=0$ string into the $N=2$ 
superstring simultaneously defines a chain of embeddings 
$N=0 \rightarrow N=1$ and $N=1 \rightarrow N=2$. As is clear from 
the relation \p{crit0}, both these intermediate embeddings do not include 
the corresponding critical ones as particular cases, in contrast 
to the resulting $N=0 \rightarrow N=2$ embedding. 

Besides two cases already considered,
there exist three other subalgebras of $N=2$
SCA which include Virasoro stress tensor $T$ :
\be
H_1=\{ T,\tilde{J} \}\; , \quad
H_2= \{ T,\GB \}\;, \quad 
H_3 =\{ T,\GB,\tilde{J} \}
\; .\nonumber
\ee
They all can be equally chosen as the linearly realized subalgebras.
The corresponding expressions for the currents, 
both on the classical and quantum level, as well as the relations 
between central charges, are presented in the next Subsections. 
In all cases we observe a remarkable matching between the critical central 
charges of $N=2$ SCA and its linearly realized subalgebras. So, once again, 
these cases admit a nice interpretation as embeddings of some 
strings into the $N=2$ superstring.

\subsection{ Stability subalgebra $H=\{ T_m,\tilde{J}_m\}$}                 
This case corresponds to the embedding of the bosonic string with
additional local $U(1)$ symmetry into the $N=2$ superstring. 

A coset element reads
\be \label{coset1}
g=e^{\frac{1}{2\pi i}\oint dz \xb(z) G(z)}
e^{\frac{1}{2\pi i}\oint dz \xi(z) \GB(z)}\;.
\ee

The currents are given by the following expressions 
\begin{eqnarray}
T&=&\frac{3}{2}\xb'\eta+\frac{1}{2}\xb \eta'+\frac{3}{2}\xi'\eb+
\frac{1}{2}\xi\eb'+T_m\;, \nn
\tilde{J}&=&-\xb\eta+\xi\eb+\tilde{J}_m\;, \nn
G&=&-\eta\;,\nn
\GB&=&-\eb+\xb\xb'\eta+\xb'\xi\eb+2\xb\xi'\eb+\xb\xi\eb'+\xb T_m+
\xb'J_m+\frac{1}{2}\xb J_m^{'}+\frac{c_m}{6}\xb''\;. \label{ccur1}
\end{eqnarray}

In order to derive quantum version we have to redefine the central charge
in the  OPEs of the subalgebra  $H=\{ T_m^q,\tilde{J}_m^q\}$
\begin{eqnarray}
T_m^q(z)T_m^q(w)&\sim&\frac{c_m/2}{(z-w)^4}+\frac{2T_m^q(w)}{(z-w)^2}+
\frac{T_m^q{}'(w)}{z-w}\;,\nn
T_m^q(z)\tilde{J}_m^q(w)&\sim&\frac{\tilde{J}_m^q(w)}{(z-w)^2}+
\frac{\tilde{J}_m^q{}'(w)}{z-w}\;,\nn
\tilde{J}_m^q(z)\tilde{J}_m^q(w)&\sim&\frac{(c_m-28)/3}{(z-w)^2}\;. 
 \label{h1}
\end{eqnarray}

The quantum correction arises only for the current $\GB$:
\be
\GB^q = \GB-\frac{11}{3}\xb''.
\ee

Now one can check that the currents 
\begin{eqnarray}
T^q&=&\frac{3}{2}\xb'\eta+\frac{1}{2}\xb \eta'+\frac{3}{2}\xi'\eb+
\frac{1}{2}\xi\eb'+T_m^q\;, \nn
\tilde{J}^q&=&-\xb\eta+\xi\eb+\tilde{J}_m^q\;, \nn
G^q&=&-\eta\;,\nn
\GB^q&=&-\eb+\xb\xb'\eta+\xb'\xi\eb+2\xb\xi'\eb+\xb\xi\eb'+\xb T_m^q+
\xb'\tilde{J}_m^q+\frac{1}{2}\xb \tilde{J}_m^q{}'+\frac{c_m-22}{6}\xb'' 
  \label{qcur1}
\end{eqnarray}
form closed quantum $N=2$ SCA \p{n2sca} with 
\be \label{crit1}
c_{N=2}=c_m-22.
\ee

To find the critical dimension for the bosonic string with additional
$U(1)$ symmetry, let us construct the BRST operator for the algebra \p{h1}.
It can be written in the form
\be
Q = \frac{1}{2\pi i}\oint dz \ [c (T_m^q+\frac{1}{2}T_{gh})+
a(\tilde{J}_m^q+\frac{1}{2}\tilde{J}_{gh})] \label{brst1} ,
\ee
where
\be
T_{gh}=cb'+2 c'b+a's \; , \; \tilde{J}_{gh} = cs'+c's \; ,
\ee
and the ghosts-anti-ghost pairs $(c,b)$, $(a,s)$ 
correspond to the $\{ T_m^q, J_m^q \}$ currents.
Condition for the nilpotency of the BRST operator \p{brst1} is $c_m=28$
which gives the correct value for the critical central charge 
of $N=2$ superstring $c_{N=2}=6$.

It is interesting
to note that the extended bosonic algebra $H=\{ T_m,\tilde{J}_m\}$ 
with the critical $c_m = 28$ naturally comes out as the algebra of 
constraints of the interacting system of string and massless particle
moving in the space with two timelike dimensions \cite{BK}. It 
was also discussed in \cite{far} in the context of `universal
string theory' and $F$-theory \cite{Vafa}. The embedding of the string 
associated with $H$ into the $N=1$ superstring possessing some 
extra symmetry ($N=1$ extension of $U(1)$ symmetry) was constructed. 
The above relations yield an alternative critical embedding of the same 
bosonic string, this time into the $N=2$ superstring. 

\subsection{ Stability subalgebra $H=\{ T_m,\GB_m \}$}
This case describes the embedding of some string which has, besides 
the Virasoro symmetry, an additional local supersymmetry generated 
by the Grassmann-odd current $\GB_m$. 

The coset element is
\begin{eqnarray}
&&g=e^{\frac{1}{2\pi i}\oint dz \xb(z)G(z)}
e^{\frac{1}{2\pi i}\oint dz \phi(z)\tilde{J}(z)} \;.
\end{eqnarray}
The expressions for currents read
\begin{eqnarray}
T&=&\frac{3}{2}\xb'\eta+\frac{1}{2}\xb \eta'+\phi'\mu +T_m\;,\nn
J&=&-\mu -\xb\eta-\frac{c_m}{6}\phi'\;,\nn
G&=&-\eta\;,\nn
\GB&=&\xb\xb'\eta+\xb\phi'\mu-\xb'\mu-\frac{1}{2}\xb\mu'+\xb T_m+
e^\phi\GB_m+\frac{c_m}{12}(2\xb''-2\xb'\phi'-\xb\phi'')\;.
\end{eqnarray}
The corresponding quantum expressions are as follows:
\begin{eqnarray}  \label{h3q}
T^q&=&T-\phi''\;,\nn
J^q&=&J+2 \phi'\;,\nn
G^q&=&G\;,\nn
\GB^q&=&\GB-\frac{3}{2}\xb''+2 \xb'\phi'\;.
\end{eqnarray}
The currents \p{h3q}
form quantum $N=2$ SCA with 
\be \label{crit3}
c_{N=2}=c_m-9\;.
\ee

BRST operator for the subalgebra $H=\{ T_m^q,\GB_m^q \}$, 
\begin{eqnarray}
T_m^q(z)T_m^q(w)&\sim&\frac{c_m/2}{(z-w)^4}+\frac{2T_m^q(w)}{(z-w)^2}+
\frac{T_m^q {}'(w)}{z-w}\;,\nn
T_m^q(z)\GB_m^q(w)&\sim&\frac{3/2\GB_m^q(w)}{(z-w)^2}+
\frac{\GB_m^q{}'(w)}{z-w}\;, \label{hh3}
\end{eqnarray}
has the following form
\be  \label{brst3}
Q = \frac{1}{2\pi i}\oint dz \ [c (T_m^q+\frac{1}{2}T_{gh})+
\alpha(\GB_m^q+\frac{1}{2}\GB_{gh})]\;,
\ee
with
\begin{eqnarray}
T_{gh}&=&cb'+2 c'b-\frac{1}{2}\alpha \beta'-
\frac{3}{2}\alpha' \beta\;, \nn
\GB_{gh}&=&\beta'c+\frac{3}{2}\beta c'\;.
\end{eqnarray}

Condition for the nilpotency of the BRST operator \p{brst3} is 
$c_m=15$, which again gives rise through \p{crit3} to the critical
central charge of the $N=2$ superstring, 
$c_{N=2} = 6$. A natural $2D$ field theory realization of the 
fermionic string associated with $H$ of the present example 
is on 10 bosonic and 10 fermionic fields, thus implying the bosonic 
critical dimension $d_m = 10$.

\subsection{ Stability subalgebra $H=\{ T_m,\GB_m,\tilde{J}_m\}$}
This case corresponds to an embedding of some string with
Grassmann-odd current $\GB_m$ and additional local $U(1)$ symmetry into 
the $N=2$ superstring. The embedded string is a `hybrid' of the strings 
associated with the stability subalgebras of two previous examples.

The OPEs of the stability subalgebra read 
\begin{eqnarray}
T_m(z)T_m(w)&\sim&\frac{c_m/2}{(z-w)^4}+\frac{2T_m(w)}{(z-w)^2}+
\frac{T_m'(w)}{z-w}\;,\;
T_m(z)\tilde{J}_m(w) \sim \frac{\tilde{J}_m(w)}{(z-w)^2}+
\frac{\tilde{J}_m'(w)}{z-w}\;,\nn
T_m(z)\GB_m(w)&\sim&\frac{3/2\GB_m(w)}{(z-w)^2}+
\frac{\GB'_m(w)}{z-w}\; , \;
\tilde{J}_m(z)\GB_m(w) \sim -\frac{\GB_m(w)}{(z-w)}\;,\nn
\tilde{J}_m(z)\tilde{J}_m(w)&\sim& \frac{c_m/3}{(z-w)^2}\;. \label{h2c}
\end{eqnarray}

The relevant coset element is defined by 
\begin{eqnarray}
&&g=e^{\frac{1}{2\pi i}\oint dz \xb(z)G(z)}\;.
\end{eqnarray}

The currents are given by the expressions 
\begin{eqnarray}
T&=&\frac{3}{2}\xb'\eta+\frac{1}{2}\xb \eta'+T_m\;,\nn
\tilde{J}&=&-\xb\eta+\tilde{J}_m\;,\nn
G&=&-\eta\;,\nn
\GB&=&\xb\xb'\eta+\xb T_m+
\xb'\tilde{J}_m+\frac{1}{2}\xb \tilde{J}_m^{'}+\GB_m+\frac{c_m}{6}\xb''\;.
\end{eqnarray}

After redefining OPEs in the subalgebra \p{h2c} as follows
\begin{eqnarray}
T_m^q(z)T_m^q(w)&\sim&\frac{c_m/2}{(z-w)^4}+\frac{2T_m^q(w)}{(z-w)^2}+
\frac{T_m^q {}'(w)}{z-w}\;,\nn
T_m^q(z)\tilde{J}_m^q(w)&\sim&\frac{-2}{(z-w)^3}+ 
\frac{\tilde{J}_m^q(w)}{(z-w)^2}+
\frac{\tilde{J}_m^q{}'(w)}{z-w}\;,\nn
T_m^q(z)\GB_m^q(w)&\sim&\frac{3/2\GB_m^q(w)}{(z-w)^2}+
\frac{\GB_m^q{}'(w)}{z-w}\; , \;
\tilde{J}_m^q(z)\GB_m^q(w) \sim -\frac{\GB_m^q(w)}{(z-w)}\;,\nn
\tilde{J}_m^q(z)\tilde{J}_m^q(w)&\sim&
+\frac{(c_m-14)/3}{(z-w)^2}\;, \label{h2q}
\end{eqnarray}
we obtain the quantum correction again only for $\GB$:
\begin{eqnarray}
\GB_q&=&\GB-\frac{11}{6}\xb''\;.
\end{eqnarray}

One can check that  the currents
\begin{eqnarray}
T^q&=&\frac{3}{2}\xb'\eta+\frac{1}{2}\xb \eta'+T_m^q\;,\nn
\tilde{J}^q&=&-\xb\eta+\tilde{J}_m^q\;,\nn
G^q&=&-\eta\;,\nn
\GB^q&=&\xb\xb'\eta+\xb T_m^q+
\xb'\tilde{J}_m^q+\frac{1}{2}\xb \tilde{J}'_m+
\GB_m^q+\frac{c_m-11}{6}\xb''
\end{eqnarray}
span a quantum $N=2$ SCA  with 
\be \label{crit2}
c_{N=2}=c_m-11\;.
\ee

The BRST operator for the subalgebra \p{h2q} is
\be  \label{brst2}
Q=\frac{1}{2\pi i}\oint dz \ [c (T_m^q+\frac{1}{2}T_{gh})+
a( \tilde{J}_m^q+\frac{1}{2} \tilde{J}_{gh})+
\alpha(\GB_m^q+\frac{1}{2}\GB_{gh})]\;,
\ee
where
\begin{eqnarray}
T_{gh}&=&cb'+2 c'b+a's-\frac{1}{2}\alpha \beta'-
\frac{3}{2}\alpha' \beta\;, \nn
\stackrel{\sim}{J}_{gh}&=&cs'+c's-\alpha \beta\;,\nn
\GB_{gh}&=&\beta'c+\frac{3}{2}\beta c'-\beta a \;
\end{eqnarray} 
and the ghost-anti-ghost pairs $(c,b)$, $(a,s)$, $(\alpha,\beta )$ 
correspond to the $\{ T_m^q, \tilde{J}_m^q ,\GB_m^q\}$ currents.

The nilpotency of the BRST operator \p{brst2} is 
achieved with $c_m=17$, 
that again yields, via eq. \p{crit2},  just the critical value for 
the central charge of $N=2$ string $c_{N=2} =6$. In accord with 
the reasoning  
at the end of previous Subsection, the critical bosonic dimension of 
the string associated with the given choice of $H$ is expected to be 
$d_m = 12$. So this string might bear a tight relation to 
the hypothetical $F$-theory. 

\subsection{Stability subalgebra $H=\{T_m, G_2\}$ and 
$N=0 \rightarrow N=1$ \protect{\\}
embeddings}

In order to show the universality of our method and reveal 
a correspondence with the embeddings of $N=0$ string into the $N=1$
superstring in the approach of refs. \cite{K, McA}, in this 
last Subsection we consider an embedding 
of the $N=1$ superstring into
the $N=2$ one in the component formalism.  
We use the real fermionic currents $G_1$ and $G_2$ satisfying 
the algebra \p{algebra}, and place them, respectively, into the coset and 
the stability subalgebra. 

We start from the coset element
\begin{eqnarray}\label{cs1}
&&g=e^{\frac{1}{2\pi i}\oint dz \xi(z)G_1(z)}
e^{\frac{1}{2\pi i}\oint dz \phi(z)J(z)}\;,
\end{eqnarray}
and find the following expressions for the $N=2$ SCA currents
\begin{eqnarray} \label{class}
T&=&\frac{3}{2}\xi'\eta+\frac{1}{2}\xi \eta'+\phi'\mu +T_m\;,\nn
J&=&-\mu -\xi\eta\tan\phi-\frac{1}{2}\xi\xi'\mu-\xi\sec\phi G_{2m}
+\frac{c_m}{12}(2\phi'+\xi\xi'\phi'+\xi\xi''\tan\phi)\;,\nn
G_1&=&-\eta-\frac{1}{2}\xi\xi'\eta-\frac{1}{2}\xi\phi'\mu
-\frac{1}{2}\xi T_m-\frac{c_m}{12}(\xi''-\frac{1}{8}\xi\xi'\xi'')\;,\nn
\label{222}G_2&=&(1-\frac{1}{2}\xi\xi')\eta\tan\phi+
\xi'\mu+\frac{1}{2}\xi\mu'-\frac{1}{2}\xi\phi'\mu\tan\phi
-\frac{1}{2}\xi\tan\phi T_m \nn
&&+(1-\frac{1}{4}\xi\xi')\sec\phi G_{2m}
-\frac{c_m}{12} (2 \phi'\xi'+\xi\phi''+\xi''\tan\phi
-\frac{3}{8}\xi\xi'\xi''\tan\phi)
\end{eqnarray}
in terms of the real Nambu-Goldstone currents $\eta(z), \xi(z)$ and their 
conjugates $\mu(z), \phi(z)$:
\begin{eqnarray}
\eta(z)\xi(w)\sim\frac{1}{z-w},\;\; \mu(z)\phi(w)\sim\frac{1}{z-w}\;.
\end{eqnarray}

The corresponding quantum expressions are
\begin{eqnarray}\label{qua}
T^q&=&T-\frac{1}{4}(\xi''\xi'+\xi'''\xi)-(\phi'\tan\phi)'\;,\nn
J^q&=&J+\frac{1}{4} \phi'\xi'\xi+\frac{5}{8}\xi''\xi\tan\phi+
    \frac{1}{2}(\tan^2\phi-3)\phi'\;,\nn
G_{1}^q&=&G_1+\frac{3}{4}\xi''+\frac{1}{4} \xi''\xi'\xi-
\frac{3}{4} \xi'\phi'\tan\phi+\frac{1}{8}\xi(\phi'\tan\phi)'\;,\nn
G_{2}^q&=&G_2+\frac{1}{16}\xi''(12+5\xi'\xi)\tan\phi+
 \frac{1}{4}\xi'\phi'(5-\sec^2\phi) \nn
 &&+\frac{1}{8}\xi \left (\phi''+\phi''\sec^2 \phi+
3 \phi'\phi'\tan\phi \sec^2\phi \right )\;.
\end{eqnarray}

Let us now consider the embedding of $N=0$ string into $N=1$ string 
within our approach and its relationship with the 
$N=1 \rightarrow N=2$ embedding just presented. 
 
We choose the $N=1$ SCA formed by the currents $T, G_1$. 
Taking as a stability
subalgebra $H=\{T_m\}$ and parametrizing the coset space as
\be\label{cs2}
g=e^{\frac{1}{2\pi i}\oint dz \xi(z)G_1(z)}\;,
\ee
we find the following expressions for the classical currents
\begin{eqnarray}
T&=&\frac{3}{2}\xi'\eta+\frac{1}{2}\xi \eta'+T_m\;,\nn
G_1&=&-\eta-\frac{1}{2}\xi\xi'\eta
-\frac{1}{2}\xi T_m-\frac{c_m}{12}(\xi''-\frac{1}{8}\xi\xi'\xi'')\;.
\end{eqnarray}
After the redefinitions
\be\label{red}
G_1=\frac{1}{\sqrt{2}}\tilde G\;, \; \xi=-\sqrt{2}\tilde \xi\;, \;
\eta =-\sqrt{2} \tilde{\eta}\;,
\ee 
the currents $T$ and $\tilde G$ precisely coincide with those deduced  
by McArthur \cite{McA}. Coincidence with the result of 
Kunitomo \cite{K} at the classical level can be further achieved 
by decoupling the matter and putting the central charge equal 
to zero~\footnote{In ref. \cite{K} the classical versions of $N=1$ 
SCA and Virasoro algebra were assumed to have a zero central charge. 
Actually, it is not necessary to require this: central charges can 
be switched on already at the classical level through, e.g., 
Feigin-Fuks terms.}.

Total expressions for the currents contain quantum corrections:
\begin{eqnarray}\label{bv}
\hat T^q&=&T_m+\frac{3}{2}\xi'\eta+\frac{1}{2}\xi \eta'
-\frac{1}{4}(\xi''\xi'+\xi'''\xi)\;,\nn
\hat G_{1}^q&=&-\eta-\frac{1}{2}\xi\xi'\eta
-\frac{1}{2}\xi T_m-\frac{c_m-11}{12}\xi''+ 
\frac{c_m-26}{96}\xi\xi'\xi''\;.\nn
\end{eqnarray}
The redefinitions \p{red} bring these expressions 
into those given by Berkovits and Vafa \cite{BV} and Berkovits 
and Ohta \cite{BO}.

Let us compare $\hat T^q$, $\hat G^q_1$ with the currents 
$T^q$, $G^q_1$ from the $N=2$ set \p{qua}. They also form 
$N=1$ SCA as a subalgebra of $N=2$ SCA and so are expected to 
admit a truncation to the expressions \p{bv}.

Despite the fact that the coset space \p{cs2} formally can be 
derived from the coset space \p{cs1} in the limit $\phi=0$, the 
relation between the corresponding quantum currents of 
$N=1$ SCA is not so simple due to the presence of conjugate variable
$\mu(z)$ $$\mu(z)\phi(w)\sim\frac{1}{z-w}$$
in the expressions \p{222}. The currents 
$T^q$ and $G^q_1$ can be brought into the following form
\begin{eqnarray}
T^q&=&(T_m+\phi'\mu-(\phi'\tan\phi)')+\frac{3}{2}\xi'\eta+
\frac{1}{2}\xi \eta'
-\frac{1}{4}(\xi''\xi'+\xi'''\xi)\;,\nn
G_{1}^q&=&-\eta-\frac{1}{2}\xi\xi'\eta
-\frac{1}{2}\xi( T_m+\phi'\mu-(\phi'\tan\phi)')
-\frac{c_m-9}{12}\xi''+\frac{c_m-24}{96}\xi\xi'\xi''\nn
 \label{addterm}
&&-\frac{3}{4} \xi'\phi'\tan\phi-\frac{3}{8}\xi(\phi'\tan\phi)'\;.
\end{eqnarray}
We observe that after passing to 
$\tilde T_m= T_m+\phi'\mu-(\phi'\tan\phi)'$,
the currents $T^q$ and $G_{1}^q$ coincide 
with those given by eqs. \p{bv}, up to the change $T_m \rightarrow 
\tilde T_m$ and the presence of two extra terms 
in $G_1^q$: 
\begin{eqnarray}
T^q &=& \hat T^q\;, \quad
G_1^q = \hat G^q_1 + \delta G_{1}^q\;, \nn 
\delta G_{1}^q &=& -\frac{3}{4} \xi'\phi'\tan\phi-
\frac{3}{8}\xi(\phi'\tan\phi)'\;.
\end{eqnarray}
It is easy to check that the addition $\delta G_1^q$ possesses the
following OPEs
\be \label{null}
T^q \;\delta G_{1}^q \sim 0,\; 
G_{1}^q \;\delta G_{1}^q+\delta G_{1}^q \;G_{1}^q \sim 0,\;
\delta G_{1}^q\; \delta G_{1}^q \sim 0\;.
\ee
Though the presence of this term in $G_{1}^q$ is
absolutely necessary for the closure of $N=2$ superconformal algebra at 
the quantum level, it is
unessential from the standpoint of its $N=1$ subalgebra 
formed by $T^q$ and $G_{1}^q$ (as follows from \p{null}, it is a 
`null field' with respect to this $N=1$ SCA). If one does 
not care about the whole $N=2$ SCA, then, as a consequence of 
the relations
\p{null}, there is a one-parameter freedom in the definition
of $G_{1}^q$:
\be
G_{1}^q(a) = \hat G_{1}^q + a\delta G_{1}^q. \label{freedom}
\ee
It means that the last term can be consistently omitted from $G_{1}^q$. 
Then the resulting expressions for the $N=1$ SCA currents coincide 
with the expressions \p{bv} in which $T_m$ is replaced by ${\tilde T}_m$. 
The presence of extra terms in $\tilde T_m$ also leads to the shift of its 
central charge to $\tilde{c}_m = c_m+2$. Thus, the final expressions 
for the currents $T^q, G_{1}^q$ reduced in this way coincide with 
the expressions \p{bv} in which $T_m, c_m$ are replaced by ${\tilde T}_m, 
\tilde{c}_m$.

We end this Section with the following comment. At any value of 
$a$ in \p{freedom} the currents $T^q$, $G_{1}^q(a)$ generate $N=1$ SCA with 
the standard relation between the $N=1$ and $N=0$ central charges 
$c_{N=1} = \tilde c_m - 11$. But only at $a=0$ one gets a minimal 
realization solely in terms of $\eta$, $\xi$ and $\tilde T_m$ that can be 
interpreted in the language of embedding of $N=0$ string into the $N=1$ 
superstring. 

\setcounter{equation}0
\section{Linearization of $W_3^{(2)}$ as a string embedding}

In this part of the article we demonstrate how the described
techniques work in the case of nonlinear algebras. As an example we take
a $W_3^{(2)}$ algebra. 
The spin content of currents of this algebra
$2, \frac{3}{2}, \frac{3}{2}, 1$
is the same as in $N=2$ SCA. The
basic difference between these two  algebras is the Grassmann parity of 
the spin $\frac{3}{2}$ currents. In the case of $W_3^{(2)}$ they are bosonic
and this fact leads to the appearance of nonlinear terms in the OPEs of
the algebra
\begin{eqnarray}
T(z)T(w)&\sim&\frac{c/2}{(z-w)^4}+\frac{2T(w)}{(z-w)^2}+
\frac{T'(w)}{z-w}\;, \nn
T(z)J(w)& \sim& \frac{J(w)}{(z-w)^2}+\frac{J'(w)}{z-w}\;, \quad
T(z)G^{\pm}(w)\sim\frac{3/2 \; G^{\pm}(w)}{(z-w)^2}+
\frac{G'^{\pm}(w)}{z-w}\;, \nn
J(z)J(w)&\sim&-\frac{c/9}{(z-w)^2}\;,\quad
J(z)G^{\pm}(w) \sim \pm\frac{G^{\pm}(w)}{(z-w)}\;, \quad
G^{\pm}(z)G^{\pm}(w)\sim 0 \;, \nn  
G^{+}(z)G^{-}(w)&\sim&\frac{-c/3}{(z-w)^3}+\frac{3J(w)}{(z-w)^2}-
\frac{T(w)+18/c \; J^2(w)-3/2 \; J'(w)}{z-w}\;. \label{OPEW}
\end{eqnarray}

The algebra \p{OPEW} is nonlinear because there is a quadratic (in $J$)
term on the right-hand side of the OPE $G^+G^-$. One way to 
construct a coset realization of this algebra along the lines of the
previous Sections is to convert it into 
an infinite-dimensional {\it linear} algebra using the trick proposed 
in \cite{IKP}. It consists in treating all the composite currents 
(including $J^2$) as some new independent currents. We use  
here another, more economic version of this approach which will allow us 
to deal only with the original finite set of $W_3^{(2)}$ currents at 
each step of calculations. 

We choose as the stability subalgebra 
the maximal linear subalgebra of $W_3^{(2)}$, that is the set 
\be \label{HW}
H = \{G^+, J, T, c\}, 
\ee
as well as all the composite currents constructed out of it, like 
$J^2$, $JT$, $cJ$, $cT$, ... . For consistency, 
we need to treat the central charge of the whole algebra also as an 
independent current and to include it from the beginning into the set 
\p{HW} on equal footing with other currents.  

The remaining current $G^-$ is placed into the coset. 
Respectively, we will interpret the quantity 
\begin{equation}           \label{k1}
K=e^{\frac{1}{2\pi i}\oint dz \xi(z) G^-(z)}
\end{equation}
as a representative of this coset. We will try to realize the 
whole $W_3^{(2)}$ as left shifts of the coset current $\xi(z)$ making use 
of the general formalism described in Sect. 2. An essentially new feature  
compared to the case of linear algebras will be the appearance of 
composite currents in the induced elements $h$ on the right hand side of 
the general relation (2.3) specialized to the present case. We will give 
a self-consistent explicit prescription of how to treat such objects 
with preserving the original algebraic structure.    

Calculating as above the left action of the infinitesimal element
\begin{equation}
\alpha =\frac{1}{2\pi i}\oint dz \{\alpha_{+}(z)G^{+}(z)+ 
\alpha_{-}(z)G^{-}(z)+
\beta(z)J(z)+a(z)T(z)\}\;,
\end{equation}
on \p{k1}, we find the variation of $\xi(x)$
\begin{equation}\label{DELTA}
\delta\xi=\alpha_{-}+\alpha_{+}\xi\xi'-
\alpha'_{+}\xi^2-\beta\xi+\frac{1}{2}
a'\xi-a\xi' \; ,
\end{equation}
and induced infinitesimal element of the subalgebra to the right of $K$
\begin{eqnarray}\label{H32}
\delta h&=& \frac{1}{2\pi i}\oint dz \{\beta J+a T+\alpha_{+}G^{+}-
\frac{c}{6}
\alpha_{+}
\xi''- 3\alpha_{+}\xi'J \nn
&&-\; \frac{3}{2}\alpha_{+}\xi J'-\alpha_{+}\xi T-
\frac{18}{c}\alpha_{+}\xi J^2+\frac{18}{c}\alpha_{+}\xi^2JG^{-}-
\frac{6}{c}\alpha_{+}\xi^3G^{-}G^{-}\}\;.
\end{eqnarray}

Since the currents $J$, $T$ and $G^-$ form a linear subalgebra in 
$W_3^{(2)}$, no problems occur while obtaining the expressions for them 
within the coset realizations approach. Just like in the cases 
considered in the previous Section, one should replace, in the 
relevant terms in \p{H32}, the involved stability subalgebra 
generators $J$, $T$, $c$ 
by their `matter' representation $J_m$, $T_m$, $c_m$ ($c_m$ is assumed to 
be a constant, and this is why we are allowed to divide by $c$ in 
\p{H32}). Also, one introduces the current $p(z)$ canonically  conjugated 
to $\xi(z)$
\begin{equation}
p(z)\xi(w)\sim\frac{1}{z-w}\;.
\end{equation}
As a result, the following expressions for $J$, $T$ and $G^-$ can be 
extracted from eqs. \p{DELTA} and \p{H32} 
\begin{eqnarray}
J = \xi p+J_m\;, \quad 
T = \frac{3}{2}\xi'p+\frac{1}{2}\xi p'+T_m\;,\quad
G^{-} = -p\;.\label{JTGmin}
\end{eqnarray}

A new characteristic feature of \p{H32} is the appearance of the composite 
currents  $J^2$, $JG^-$, $G^-G^-$ in the expression for the induced 
element of the subalgebra corresponding to the left shift by 
the generator $G^+$ (the coefficient before the parameter $\alpha_+$ in 
the r.h.s of \p{H32}). This comes about just due to the presence of 
nonlinear term in the last of OPEs \p{OPEW}, so one can expect that 
such a phenomenon is typical for nonlinear algebras. 
This is a crucial difference compared to the previously considered 
linear case, and one should give a recipe of how to treat such composite 
currents within the nonlinear realizations formalism.

Once again, in accord with the general 
prescriptions of Sect. 2, we should replace the stability subalgebra 
generators by their `matter' representation $\{G^+_m, J_m, T_m, c_m\}$ 
wherever they appear, on their own or as building blocks of composite 
currents. What concerns the composite currents 
containing $G^-$, namely $JG^-$ and $G^-G^-$,  
a natural idea is to replace the current $G^-$ in them by its already 
found coset expression \p{JTGmin}. In this way we get 
\begin{eqnarray}
G^{+}&=&G^{+}_m-3\xi\xi'p-\xi^{2}p'-\frac{c}{6}\xi''-3\xi'J_m-\frac{3}{2}
\xi J'_m-\xi T_m 
- \frac{18}{c}\xi J_m^2 \nn
&& - \;\frac{18}{c}\xi^2 J_mp-\frac{6}{c}\xi^3p^2\;.  
\label{lin} 
\end{eqnarray}
It is straightforward to verify that the currents \p{JTGmin}, \p{lin} 
satisfy just the OPEs \p{OPEW} with $c= c_m$. 

A consistency check of our procedure of deriving the coset 
representation for the $W_3^{(2)}$ currents goes as follows. We act 
on the coset element \p{k1} from the left by composite currents 
constructed out of the original abstract currents $J$, $T$, $G^{\pm}$ 
and find their coset realization by pulling them through 
\p{k1} with the use of OPEs \p{OPEW} and finally making 
the changes $\{J, T, G^+, c\} \rightarrow \{J_m, T_m, G_m^+, c_m\}$ 
and $G^- \rightarrow -p$. We explicitly found such a realization for 
the composite currents $J^2$, $G^-G^-$, $JG^+$. In all cases, 
the resulting expressions are given by the appropriate products of 
the currents \p{JTGmin} and \p{lin}. 

Note that this modified nonlinear (coset) realizations scheme seems to 
work, with minor further modifications, in the case of other nonlinear 
algebras too.

Similarly to the $N=2$ SCA realizations constructed in Sect. 3, the 
formulas \p{JTGmin}, \p{lin} give a realization of the $W_3^{(2)}$ 
generators in terms 
of Nambu-Goldstone fields $\xi$, $p$ and a closed set of `matter' 
currents $J_m, T_m, G^+_m$. It can be checked that these 
expressions coincide with the classical limit of
the expressions for the $W_3^{(2)}$ currents obtained in \cite{KS} 
within the procedure of conformal linearization of $W_3^{(2)}$. 
The currents 
$J_m, T_m, G^+_m$ together with the `ghost-antighost' pair $\xi, p$ 
form just the classical version of the linearizing algebra 
for $W_3^{(2)}$. Thus at the classical level the `linearization' of 
$W_3^{(2)}$ amounts to constructing its above coset realization, with 
the Nambu-Goldstone field $\xi$ (and its conjugate momenta $p$) playing the 
role of additional currents which allow to linearize $W_3^{(2)}$ 
in the spirit of ref. \cite{KS}. One encounters here a surprising 
situation when two nonlinearities (the intrinsic nonlinearity of 
$W_3^{(2)}$ algebra and the nonlinearity of the coset realization procedure) 
`interfere' to yield a linearity in the end. It would be interesting to 
apply the same techniques to construct linearizing algebras for those 
$W$ algebras for which this construction is still lacking, e.g., 
$SO(N)$ Knizhnik-Bershadsky algebra. 

It is tempting to make a step further and to wonder whether this 
linearization procedure through a nonlinear realization of $W_3^{(2)}$ 
admits an interpretation in terms of string embeddings, like nonlinear 
realizations of $N=2$ SCA constructed in the previous Section. It does! 
To see this, one needs, first of all, to pass \cite{KS} to the quantum 
counterparts of eqs. \p{JTGmin}, \p{lin} 
\begin{eqnarray} 
J^q & = & J , \quad 
T^q  =  T+\frac{9}{9-c}J^q{}' ,\quad 
G^{-}{}^q  =  G^{-} \;, \nn 
G^{+}{}^q & = & G^{+}_m -\xi T^q +\frac{18}{9-c}\xi J^q J^q +
  \frac{3(21-c)}{2(9-c)}\xi J^q{}'-\frac{18}{9-c}\xi^2 pJ^q \nn 
  & & + \;\frac{6}{9-c}\xi^3 p^2 +3 \xi\xi' p +\frac{9+c}{9-c}\xi^2 p'-
   3\left( \xi J^q-\frac{9-c}{18}\xi'\right)'\;.
\label{linqua}
\end{eqnarray}
An essential peculiarity of this case originating from the nonlinear nature 
of $W_3^{(2)}$ and having no analog in the case of $N=2$ SCA 
(and other linear conformal algebras) is that not only the central 
charges of $W_3^{(2)}$ and the subalgebra $\{ T_m, J_m, G^+_m \}$ become 
different after passing to the quantum case, but also the structure 
relations of the latter algebra are modified. Namely, its quantum OPEs 
are as follows 
\bea
T_m(z_1)T_m(z_2) & \sim &
 \frac{9+4c-c^2}{2(9-c)}\frac{1}{z_{12}^4}+
        \frac{2T_m}{z_{12}^2}+\frac{T_m'}{z_{12}} \; , \nn
T_m(z_1)G^{+}_m(z_2)  & \sim &
  \left[ \frac{3}{2}+\frac{9}{9-c} \right]\frac{G^{+}_m}{z_{12}^2}+
           \frac{G^{+}_m{}'}{z_{12}} \;, \;
T_m(z_1)J_m(z_2)  \sim 
       \frac{J_m}{z_{12}^2}+\frac{J_m'}{z_{12}} \; , \nn
J_m(z_1)J_m(z_2)  &\sim &   \frac{9-c}{9z_{12}^2}  \; , \;
J_m(z_1)G^{+}_m(z_2)  \sim  \frac{G^{+}_m}{z_{12}}\;, 
\eea
and they differ from the classical ones in that the current $G^+_m$ 
acquires an anomalous conformal dimension ${3\over 2} + {9\over 9-c}$. 
Actually, this is none other than the algebra $W^{lin}_3$ which linearizes 
another kind of nonlinear algebra, Zamolodchikov's $W_3$ algebra \cite{KS}.
Now, we wish to interpret eqs. \p{linqua} as describing an embedding of 
some string with the symmetry algebra $W^{lin}_3$ in the matter 
sector into a string associated with $W_3^{(2)}$. A consistency check 
for such an interpretation is to inquire whether 
the critical central charges of these two algebras match with each other. 
They can be parametrized in terms of a single parameter $c$ as follows 
\cite{KS}
\be 
c_{(TT)} = \frac{(7+c)c}{c - 9}\;, \; c_m = \frac{9+4c-c^2}{9-c}\;. 
\label{cwcm}
\ee
The BRST operator for $W^{lin}_3$ was constructed in \cite{boh}, it is
nilpotent 
at $c = 18$ and, accordingly, $c_m = 27$. On the other hand, as noted in 
the same paper \cite{boh}, the nilpotency condition for the BRST charge
operator 
for $W_3^{(2)}$ \cite{khvsezg} requires $c_{(TT)} = 50$. But this is 
just the 
value that arises upon substitution, into $c_{TT}$ in \p{cwcm}, of $c = 18$ 
which simultaneously produces the critical value for 
the central charge $c_m$. 
Thus the remarkable matching between critical central charges which is a
characteristic feature of embeddings in the case of linear superconformal 
algebras extends to this nonlinear case as well. This is a strong indication 
that in the present case we indeed deal with an embedding of 
string associated with the linear algebra $W^{lin}_3$ 
into the $W_3^{(2)}$ string~\footnote{In order to rigorously prove 
this conjecture, one should show that the cohomology of the BRST 
operator for $W_3^{(2)}$ in this specific realization coincides with 
that of the BRST operator for  $W^{lin}_3$.}. Note that 
an analogous relation between critical 
central charges was found in \cite{boh} while considering an embedding 
of the bosonic string (associated with the Virasoro subalgebra $T_m$) into 
the $W_3^{(2)}$ string. It would be interesting to check whether such an 
embedding can be reproduced within our nonlinear realization approach.

\setcounter{equation}0\section{Conclusions}

In this paper we presented a modified nonlinear realization 
method directly applicable to superconformal and some $W$ type 
algebras in the formulation based on OPEs or SOPEs of the 
relevant currents or supercurrents. This provides a 
systematic way of deducing the relations describing various embeddings 
of bosonic and fermionic strings. The embedded string or superstring 
(its matter sector, to be precise) always corresponds to the vacuum 
stability subalgebra of the given nonlinear realization while for the 
(super)currents of the 
embracing algebra one algorithmically gets the expressions in terms of 
the appropriate coset (super)fields, their conjugate momenta and the 
generators of the stability subalgebra. The method as it stands  
is limited to the classical algebras (although with non-zero central 
charges, as distinct from the approach of \cite{K}, say). Nonetheless,
it is straightforward 
to explicitly find the quantum corrections and to get the genuine relations 
describing the string embeddings. We reproduced in this way some known 
examples ($N=0 \rightarrow N=1$ \cite{BV}, $N=1 \rightarrow N=2$ \cite{BO}) 
and constructed new embeddings of the bosonic string and some of its
extensions into the $N=2$ string. It would be interesting to reveal 
possible physical implications of such extended strings and to identify 
their place in the modern string realm. All the 
embeddings constructed include as a particular case the corresponding 
critical embedding.

We also applied our method to an example of nonlinear 
$W$ type algebras, $W_3^{(2)}$ algebra, choosing as the stability 
subalgebra the maximal linear subalgebra of $W_3^{(2)}$. It coincides 
with the algebra $W^{lin}_3$ introduced in \cite{KS} as the linearizing 
algebra for Zamolodchikov's $W_3$. Surprisingly, our approach 
immediately leads to the relations of ref. \cite{KS} describing the 
linearization of $W_3^{(2)}$. Thus the linearization procedure 
for this particular $W$ algebra can be as well understood as the
embedding of the $W^{lin}_3$ string into the $W_3^{(2)}$ string.
The critical central charges in this case nicely match with each other,
similarly to other examples considered. An interesting problem is to 
treat a wider class of $W$ type algebras 
from the point of view presented here and to see whether 
the linearization procedure for them \cite{a1}-\cite{a3} always 
admits an interpretation in terms of proper string embeddings.      

As a prospect for further developments we mention possible 
applications of our method to other superconformal algebras, 
e.g. `small' and `large' $N=4$ SCAs. Some string embeddings related to 
these algebras were described in \cite{BO,OS}. Our method will hopefully 
allow to list all possible such embeddings by choosing various 
subalgebras of $N=4$ SCAs as the vacuum stability ones.  It would be 
extremely interesting to generalize our method to  
more complicated extended objects like $p$-branes, i.e. to learn how to 
construct nonlinear realizations of the relevant infinite-dimensional 
world-volume gauge symmetries and to describe embeddings of such 
objects in this universal language~\footnote{See a recent preprint 
\cite{far} for an attempt to apply the embedding 
ideas in the context of $F$-theory.}.        

Finally, let us remind that nonlinear realizations of 
$1D$ superconformal algebras and $W$ algebras were previously considered 
from various points of view in a number of works, e.g. in 
\cite{bi1,IKP}, \cite{ss}-\cite{ikl}. 
It is of interest to understand  
how these approaches are related with the one presented here.

\vspace{0.3cm}
\noindent {\bf Acknowledgments}

\vspace{0.2cm}
\noindent Two of us (S.K. and A.P.) are indebted to D. L\"ust and 
C. Preitschopf for hospitality at the Humboldt University in Berlin,
where a part of this work has been done.
This investigation has been supported in part by the 
Fondo Scambi Internazionali Convenzione Particellare
INFN-Dubna, the Russian Foundation of
Fundamental Research, grant RFFR 96-02-17634, joint grant RFFR-DFG
96-02-00186G, INTAS grants 93-127 and 94-2317 and
by a grant of the Dutch NWO Organization.

\end{document}